# Mist-Assisted Federated Learning for Intrusion Detection in Heterogeneous IoT Networks

Saadat Izadi
Department of Computer Engineering, Razi University, Kermanshah, Iran
s.izadi@razi.ac.ir

Shakib Komasi
Department of Computer Engineering, Razi University, Kermanshah, Iran
sh.komasi@razi.ac.ir

Ali Salimi
Department of Computer Engineering, Razi University, Kermanshah, Iran
a.salimi@stu.razi.ac.ir

Alireza Rezaei
Department of Computer Engineering, Gilan University, Gilan, Iran
a.rezaei@webmail.guilan.ac.ir

Mahmood Ahmadi*
Department of Computer Engineering, Razi University, Kermanshah, Iran
m.ahmadi@razi.ac.ir
* Corresponding author

*Abstract*— **The rapid growth of the Internet of Things (IoT) offers new opportunities but also expands the attack surface of distributed, resource-limited devices. Intrusion detection in such environments is difficult due to data heterogeneity from diverse sensing modalities and the non-IID distribution of samples across clients. Federated Learning (FL) provides a privacy-preserving alternative to centralized training, yet conventional frameworks struggle under these conditions. To address this, we propose a Mist-assisted hierarchical framework for IoT intrusion detection. The architecture spans four layers: (i) Mist, where raw data are abstracted into a unified feature space and lightweight models detect anomalies; (ii) Edge, which applies utility-based client selection; (iii) Fog, where multiple regional aggregators use FedProx to stabilize training; and (iv) Cloud, which consolidates and disseminates global models. Evaluations on the TON-IoT dataset show the framework achieves 98–99% accuracy, PR-AUC >0.97, and stable convergence under heterogeneous and large-scale settings, while maintaining efficiency and preserving privacy.**

## I. Introduction

The Internet of Things (IoT) has rapidly evolved into one of the most pervasive technological ecosystems of the digital era, enabling seamless interconnection across industrial, healthcare, transportation, and smart home domains [1]. Recent market analyses report that the number of connected IoT devices exceeded 18.8 billion in 2024 and is projected to approach 40 billion by 2030, underscoring the unprecedented scale and ubiquity of this paradigm [2]. In this context, Intrusion Detection Systems (IDSs) emerge as a cornerstone for securing IoT ecosystems. IDSs monitor network traffic, device behavior, and system logs to identify malicious patterns and anomalies [3, 4].

In recent years, Federated Learning (FL) has emerged as a promising paradigm for intrusion detection in IoT networks, offering privacy-preserving and communication-efficient training by keeping raw data localized at the devices while sharing only model updates with higher layers [5]. Despite its advantages, deploying FL in heterogeneous IoT environments remains highly challenging. IoT devices differ widely in sensing modalities, computational capabilities, and data distributions, resulting in non-IID and high-dimensional data heterogeneity that undermines model convergence and detection accuracy [6]. To mitigate these issues, several studies have proposed strategies at the Edge layer [7, 8]. While these approaches alleviate some of the adverse effects of heterogeneity, they often rely on mid-layer aggregation and overlook the fact that raw heterogeneity originates directly at the sensing layer, leaving room for further innovation.

While prior efforts at the Edge layer have partially alleviated heterogeneity, a key observation is that data heterogeneity originates directly at the sensing devices. Recent advances in on-device intelligence and TinyML demonstrate that lightweight feature extraction and anomaly detection can be performed at the device level without significant resource overhead [9-11]. This opens new opportunities for addressing heterogeneity closer to the data source, at the Mist layer, where raw and highly diverse data are first generated. By introducing feature abstraction mechanisms directly at the sensing layer, heterogeneous inputs such as environmental signals, network flows, and system logs can be mapped into a unified feature space before participating in federated training. The emergence of mist computing and the introduction of an additional layer close to the data source (data locality) within the conventional edge–fog–cloud architecture has attracted significant attention from researchers. This layer enables preliminary local processing, feature abstraction, and even early anomaly detection, while simultaneously reducing transmission overhead and privacy risks. In the latest frameworks, the mist layer is responsible for executing lightweight models (e.g., autoencoders or CNNs), transmitting only feature vectors or model parameters to higher layers [12].

In light of these challenges, a novel framework is introduced that leverages Mist-assisted federated learning for intrusion detection in heterogeneous IoT environments. The proposed architecture distributes intelligence across four layers Mist, Edge, Fog, and Cloud where data heterogeneity is mitigated at the Mist through feature abstraction, client selection is optimized at the Edge, and robust model aggregation is achieved at multiple Fog nodes before being consolidated into a global model at the Cloud. To validate this approach, experiments are conducted on the TON-IoT [13] dataset, a benchmark that integrates highly heterogeneous data sources such as network traffic, telemetry from IoT devices (e.g., fridge, thermostat, GPS tracker), and system logs. This dataset represents real-world IoT heterogeneity, making it particularly suitable for evaluating intrusion detection strategies under non-IID conditions. The main contributions of this work are:

- **Novel integration of Mist computing:** unlike most existing studies that focus primarily on Edge–Fog–Cloud architectures, our work explicitly integrates the Mist layer to perform lightweight feature abstraction directly at sensor nodes. This enables early anomaly detection, reduces communication overhead, and addresses heterogeneity in raw IoT data stream;

- **Client selection strategy in the Edge layer:** we introduce a data-quality–aware client selection mechanism in the Edge layer, which prioritizes devices based on dataset size, data quality, and computational capacity. This strategy is further combined with FedProx[1] [14] aggregation in the Fog layer to enhance convergence under non-IID data distributions;

- **Hierarchical four-layer architecture (Mist–Edge–Fog–Cloud):** we design and evaluate a novel four-tier architecture that distributes responsibilities across Mist, Edge, Fog, and Cloud layers. This hierarchy ensures low-latency detection near data sources, efficient aggregation in intermediate layers, and reliable global orchestration in the Cloud, thereby achieving both scalability and accuracy;

- **Comprehensive evaluation on real-world IoT datasets:** extensive experiments on benchmark IoT intrusion detection datasets demonstrate that our approach achieves higher detection accuracy, faster convergence, and lower communication cost than state-of-the-art federated IDS frameworks.

The organization of the paper is as follows: In Section 2, a review of related works in the field of intrusion detection in the IoT using FL is presented. The proposed approach is described in Section 3, and the evaluation results and comparison with previous approaches are discussed in Section 4. Finally, in Section 5, the conclusion and future work are presented.

## II. RELATED WORK

Intrusion detection in Internet of Things (IoT) networks is a challenging problem due to large-scale deployment, resource-constrained devices, and the inherently distributed nature of data. Extensive research has shown that machine learning [4] and deep learning (DL) methods are effective tools for IoT intrusion detection because they can extract attack patterns from network traffic and logs [15]. However, these approaches face several important challenges, including high resource consumption on constrained devices and centralized processing, increased communication overhead and cost due to raw-data transmission, privacy concerns, poor generalization under heterogeneous (non-IID) data distributions, and elevated false-positive rates [16]. To address privacy concerns and reduce the need for transmitting raw data, Federated Learning (FL) has rapidly emerged as a leading paradigm for intrusion detection systems (IDSs) in distributed IoT environments [17, 18]. Some studies highlight that federated learning (FL) preserves detection performance while maintaining data locality on devices, yet it faces persistent challenges in dealing with non-IID data, limited computational resources, and network latency [19, 20].

In the paper [21] resents a stacked-unsupervised federated learning (FL) approach for a flow-based network intrusion detection system (NIDS) on a cross-silo configuration. The approach uses a deep autoencoder and energy flow classifier in ensemble learning, outperforming traditional local learning and naive cross-evaluation. It also performs well in non-IID data silos and is recommended for generalization between heterogeneous networks. Shukla et al. [22] introduced a cloud-assisted federated learning framework that combined heterogeneous models for malware detection in IoT environments. Their experiments, conducted on Raspberry Pi devices, demonstrated that detection accuracy could be enhanced while incurring only modest computational costs. Recent research has primarily focused on enhancing the performance of intrusion detection systems (IDS) in IoT networks by leveraging multi-layer architectures mitigate challenges like data heterogeneity and non-uniform (non-IID) sample distributions. Specifically propose hierarchical Fog–Edge–Cloud architectures for IoT IDS to move processing closer to the data source and reduce cloud load [23].

Tariq et al. [24] proposed a fog–edge–enabled federated learning intrusion-detection framework for smart grids. Their method trains a Support Vector Machine collaboratively across edge devices, sharing only model parameters to preserve data privacy. Evaluated on the NSL-KDD and CICIDS2017 datasets, the approach outperformed conventional cloud-based IDS models, showing notable gains in accuracy, precision, recall, and F1 score. This work highlights the effectiveness of privacy-preserving FL for distributed smart-grid security. Qin et al. [25] introduced Hier-SFL, a three-tier client–edge–cloud framework for network-traffic classification. The method combines federated and split learning to address shifting traffic distributions, scarce labeled data, and uneven computing resources. By distributing neural-network training across devices and periodically aggregating model parameters, Hier-SFL achieves higher training efficiency and better adaptability to open, complex network environments than conventional supervised approaches. Despite these advances, existing methods still struggle with heterogeneity and non-IID data, often leading to unstable convergence and high false positives. Prior works focus mainly on edge and fog layers, overlooking the data-source level where heterogeneity arises. To address this gap, we propose a Mist-assisted hierarchical framework that tackles heterogeneity directly at the Mist layer and ensures robust, scalable, and privacy-preserving intrusion detection.

---

[1] FedProx is a widely used federated optimization algorithm introduced to improve convergence under heterogeneous and non-IID environments.

### III. PROPOSED METHOD

In this section, the problem is first defined to highlight the key challenges of intrusion detection in IoT. Then, the proposed architecture and its federated training process are described, followed by the mathematical formulation of the framework.

#### A. Problem Definition

Intrusion detection in IoT networks is confronted with structural limitations that undermine the effectiveness of both centralized and federated approaches. Intrusion detection in IoT networks is hindered by two fundamental challenges: (i) data heterogeneity across devices and (ii) the statistical non-IID nature of distributed data. The first challenge is data heterogeneity across devices. IoT environments integrate highly diverse data modalities environmental sensors generate continuous numerical streams, networking modules produce packet flows, and computing systems log textual events. Such heterogeneity in data type and dimensionality complicates the learning process, leading to unstable convergence, higher false positive rates (FPR), and reduced detection accuracy when conventional or even classical federated learning (FL) methods are applied. The second challenge is the statistical non-IID nature of IoT data. In practice, the distribution of data samples across devices is highly imbalanced; some clients predominantly store normal traffic while others capture disproportionately more attack-related samples. This statistical skewness destabilizes the federated optimization process and further amplifies model bias.

Beyond these two challenges, practical deployment also faces resource and privacy constraints. Transmitting raw, heterogeneous IoT data to the cloud is not a feasible solution, since it imposes excessive communication overheads and threatens user privacy. Existing methods have attempted to mitigate heterogeneity by applying client selection and aggregation strategies at the Edge layer; however, these mid-layer interventions fail to fully address the fact that heterogeneity originates at the data source itself. Consequently, there remains a pressing need for a framework that tackles heterogeneity directly at the sensing layer, while also ensuring robustness to non-IID data distributions, scalability across devices, and preservation of user privacy.

#### B. Proposed Architecture

To overcome these challenges, we design a four-layer architecture consisting of Mist, Edge, Fog, and Cloud layers (Fig. 1). The figure illustrates the flow of data and model updates across the four layers and highlights the specific role each layer plays in the federated training process. In the Mist layer, each IoT device locally processes its raw data through a lightweight encoder to obtain a fixed-size feature vector of dimension k (feature abstraction). This transformation ensures that heterogeneous data sources are mapped into a common feature space. It should be noted that this process mainly handles input-space heterogeneity. The statistical non-IID heterogeneity across clients is mitigated by FedProx aggregation at the Fog layer, which stabilizes training and balances updates. Each IoT device employs a shared lightweight encoder that is pre-trained at the Fog layer on a small representative dataset and later fine-tuned locally with few-shot adaptation. This approach ensures that even with a limited number of samples, Mist nodes can still perform efficient feature abstraction without overfitting. The encoder maps heterogeneous inputs into a unified latent space, while deeper layers (Fog and Cloud) handle statistical heterogeneity. In addition to sending model updates, Mist devices can also detect and report obvious anomalies directly at the local level. In the Edge layer, neither raw data nor model parameters are aggregated. Instead, the primary responsibility of this layer is client selection. Each Mist device reports metadata such as local dataset size ($n_i$), data quality indicator ($Q_i$), and participation history. The data-quality indicator $Q_i$ is computed as the ratio of valid (non-corrupted) samples and the entropy of feature distributions in the local dataset. Each node periodically reports $|D_i|$ and $Q_i$ to its corresponding Edge server. The Edge layer then calculates a composite utility score for each client: $U_i = \alpha |D_i| + \beta Q_i$, where $\alpha$ and $\beta$ are weighting factors determined empirically. Clients with higher $U_i$ values are prioritized for participation to improve stability and reduce the risk of noisy updates.

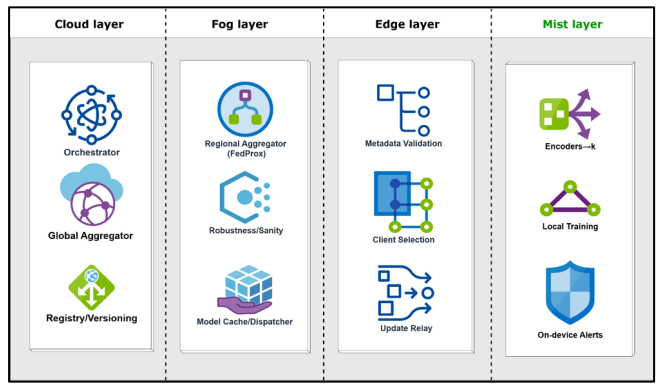

*Figure 1. Hierarchical Mist–Edge–Fog–Cloud architecture for intrusion detection. Mist devices abstract data and perform local training, Edge nodes validate metadata and select clients, Fog nodes aggregate updates with FedProx for stability, and the Cloud layer consolidates and versions the global model.*

In the Fog layer, multiple Fog nodes are deployed to aggregate updates from their respective groups of Edge nodes. Each Fog node applies the FedProx algorithm, which augments the local objective with a proximal term. This prevents local updates from drifting too far from the Fog model, thereby stabilizing training under non-IID conditions. The outcome of this process is a Fog-level model at each node. In the Cloud layer, the models from all Fog nodes are transmitted to the central server in the cloud. The Cloud performs the role of a global aggregator, consolidating the Fog models into a single global model. This global model is stored, versioned, and redistributed to the Fog, Edge, and Mist layers for the next training round, ensuring synchronization and continuity across the hierarchy.

*C. Federated Training Process*

Training proceeds in the form of federated rounds. At the beginning of each round, the Cloud sends the current global model to all Fog nodes. Each Fog forwards the model to its connected Edge nodes, which then distribute it only to the selected Mist clients. Each Mist device trains the lightweight model for one or more local epochs using its private data and computes a parameter update $\Delta w_i$. The Edge collects updates from eligible clients and passes them to its corresponding Fog node. Each Fog node aggregates the received updates using FedProx to produce an improved Fog model. These Fog models are then sent to the Cloud, where the global aggregator combines them into the final global model $w_{\text{global}}^{(t+1)}$. The updated global model is then pushed back down the hierarchy and stored in the Cloud as the new global model. This cycle repeats until the model reaches convergence.

*D. Mathematical Formulation*

Suppose the set of IoT clients is defined as $S = \{s_1, s_2, ..., s_N\}$, where each client $s_i$ holds a local dataset $\mathcal{D}_i = \{(x_{ij}, y_{ij})\}_{j=1}^{n_i}$ with $n_i$ samples. Each raw data point $x_{ij} \in \mathbb{R}^{d_i}$ lies in a feature space of dimension $d_i$ depending on the sensor type. To address data heterogeneity, the Mist layer employs a lightweight encoder $f_{\theta_i}: \mathbb{R}^{d_i} \to \mathbb{R}^k$ to transform raw data into a standardized k dimensional feature vector (1):

$$z_{ij} = f_{\theta_i}(x_{ij}) \quad (1)$$

where $\theta_i$ are the encoder parameters. On top of these abstracted features, a lightweight local model $h_{w_i}$ is trained. The local objective function of client i is defined as:

$$\mathcal{L}_i(w_i) = \frac{1}{n_i}\sum_{j=1}^{n_i} \ell(h_{w_i}(z_{ij}), y_{ij}) \quad (2)$$

where $w_i$ are the parameters of the local model, and $\ell(0)$ denotes the loss function, such as cross-entropy or reconstruction loss. At the beginning of each federated learning round t, the Edge layer collects metadata including dataset size $n_i$ and data quality indicator $Q_i$. The set of active clients is defined as:

$$\mathcal{C}_t = \{i \mid n_i \geq \tau_n, Q_i \geq \tau_q\} \quad (3)$$

where $\tau_n$ and $\tau_q$ are predefined thresholds for dataset size and quality. In the Fog layer, updates from active clients are aggregated. To address non-IID data distributions, FedProx is employed, reformulating the local optimization problem as:

$$\min_{w_i} \mathcal{L}_i(w_i) + \frac{\mu}{2} \|w_i - w_{\text{fog}}^{(t)}\|_2^p \quad (4)$$

where $w_{\text{fog}}^{(t)}$ denotes the Fog model parameters at round t, and $\mu > 0$ is the proximal coefficient ensuring stable convergence. Finally, Fog aggregates the received local updates using a weighted average:

$$w_{\text{fog}}^{(t+1)} = \sum_{i \in \mathcal{C}_t} \frac{n_i}{\sum_{j \in \mathcal{C}_t} n_j} w_i^{(t)} \quad (5)$$

These updated Fog models are transmitted to the Cloud, where the global aggregator consolidates them into the final global model $w_{\text{fog}}^{(t+1)}$. This global model is then distributed back to Fog, Edge, and Mist layers to initiate the next round of training.

## IV. EVALUATION

In the section, the parameters of the proposed approach, the dataset used, the evaluation of results, and a comparison with other studies will be discussed.

*A. Experimental setting and parameter analysis*

*1) Computing Environment:* All implementations and evaluations were conducted on a workstation running Windows 10 with an Intel® Core™ i7-7500U CPU, 12 GB RAM, and an NVIDIA GeForce 930MX GPU. Experiments were executed in the Spyder IDE using Python 3.13.3, CUDA 13.0.

*2) Local Model at Mist:* Each client trains a compact encoder–classifier: a lightweight 1D-CNN feature encoder (three Conv–BN–ReLU blocks with 3×1 kernels and stride 1, followed by max-pooling) feeding a 2-layer MLP head with dropout 0.1 and a softmax output. Cross-entropy loss is optimized via Adam (batch size 32; local epochs = 5).

*3) Federated Protocol:* Training follows synchronous rounds with FedProx aggregation at Fog nodes; the proximal coefficient is μ=0.01. We sweep clients = {10, 20, …, 100} and rounds = {50, 100, 150, 200}; in each setting, all available clients participate to isolate the effect of heterogeneity and scale. Although the experimental setup was limited to a maximum of 100 clients due to computational constraints, the four-layer Mist–Edge–Fog–Cloud architecture is designed for scalability to thousands of devices in real IoT deployments. The reduced number of clients in our experiments reflects experimental feasibility, not architectural necessity. Fog nodes aggregate client updates with FedProx and forward Fog-level models to the Cloud, where a global model is consolidated. (Utility-based selection at Edge is available but disabled during sweeps to avoid confounding factors.) Reported core settings: dataset = TON-IoT, algorithm = FedProx, μ=0.01, local epochs = 5, batch size = 32.

*4) Evaluation Metrics:* For evaluation, several complementary metrics were adopted to assess both detection effectiveness and system performance. Accuracy, along with Receiver Operating Characteristic (ROC) and Precision–Recall (PR) curves with their AUC values, was used to examine classification quality under heterogeneous and imbalanced conditions. The F1-Score distribution across training rounds provided insights into convergence stability, while model drift was analyzed as an indicator of consistency among client updates in non-IID scenarios. In addition, the distribution of experiment times was measured to capture computational overhead and scalability. Together, these metrics provide a holistic view of the robustness, stability, and efficiency of the proposed Mist–Edge–Fog–Cloud framework.

*B. Dataset*

The proposed method was evaluated using the ToN-IoT dataset [12], which is designed to collect and analyze diverse data sources from the IoT and Industrial IoT (IIoT). This dataset consists of heterogeneous data collected from various sources. The dataset was gathered with the aim of assessing the accuracy and efficiency of different cybersecurity applications leveraging artificial intelligence. To emulate device diversity and statistical skew, data are non-IID partitioned by source/type across clients (S= {10, 20, …, 100}), with an 80/20 split for local train/validation on each client. Feature abstraction at the Mist layer maps raw inputs to a fixed k-dimensional vector (k=64), ensuring a uniform representation before FL.

*C. Result*

The experimental results are presented in this section, highlighting the performance, robustness, and efficiency of the proposed framework under heterogeneous IoT conditions.

Figure 2 plots ROC curves for 10–100 clients. All curves sit well above the random baseline, with AUC = 0.814–0.856; smaller client sets (10, 30) achieve the highest AUCs, reflecting lower heterogeneity. Even at 100 clients, AUC remains >0.82, showing robustness under strong non-IID conditions. The envelopes lie close to the top-left region, which is consistent with the 98–99% final accuracy observed at our operating threshold in the accuracy plots, indicating a favorable TPR–FPR trade-off. Together with the stabilized loss across rounds, these ROC shapes confirm convergence and sustained discriminative power across participation scales.

Complementary insights are provided by the Precision–Recall (PR) curves shown in Figure 3. Figure 3 illustrates the PR curves for varying numbers of clients (10, 30, 50, 70, 100). Across all settings, the PR-AUC consistently remains above 0.97, confirming the strong resilience of the proposed framework under class-imbalanced intrusion detection scenarios. The 50-client configuration achieves the highest PR-AUC, approaching 0.99, reflecting the most balanced trade-off between precision and recall. Similarly, the 10-client and 30-client cases maintain PR-AUC values around 0.98, benefiting from reduced heterogeneity. Although performance slightly decreases as the number of clients grows to 70 and 100, their PR-AUC values remain robust at ≥ 0.97, which still indicates near-optimal detection capability. These results highlight that the proposed hierarchical Mist–Edge–Fog–Cloud architecture sustains exceptionally high precision and recall, even when scaling to larger federated deployments. Importantly, the curves converge smoothly towards the top-right corner, emphasizing reliable intrusion detection with minimal false negatives and false positives.

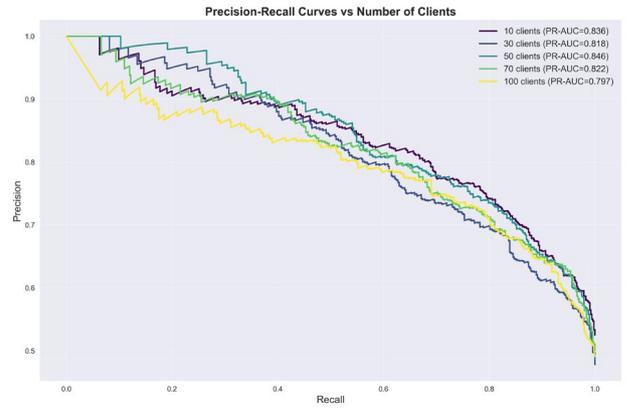

*Figure 3. Precision–Recall curves for different client counts (10, 30, 50, 70, 100)*

The distribution of accuracy across client counts is shown in Figure 4. For smaller client counts (10–30), accuracy is tightly distributed around very high values (≈ 0.98), indicating both stability and consistency. As the number of clients grows beyond 50–70, accuracy remains relatively strong (> 0.97) but shows increased variance, highlighting the effects of heterogeneity and non-IID data. For larger client counts (90–100), the distribution shifts downward with medians around 0.93–0.94, accompanied by wider interquartile ranges and more outliers. This confirms that while the proposed framework is resilient, excessive client participation introduces instability, likely due to diverse data quality and communication overhead.

Figure 5 reveals how additional training rounds impact the F1-Score distribution. At 50 rounds, the F1-Score median is around 0.90 with noticeable variability, showing that early rounds are insufficient for consistent convergence. With 100 rounds, both the median and upper quartile improve (~0.93), and variability decreases. At 150 rounds, performance reaches higher stability, with many instances exceeding 0.94. Finally, at 200 rounds, the model consistently achieves strong F1-Scores (≈ 0.95) with reduced spread, demonstrating robust convergence. This trend confirms that extended training rounds significantly enhance detection stability and precision, ensuring the model generalizes better across heterogeneous IoT clients.

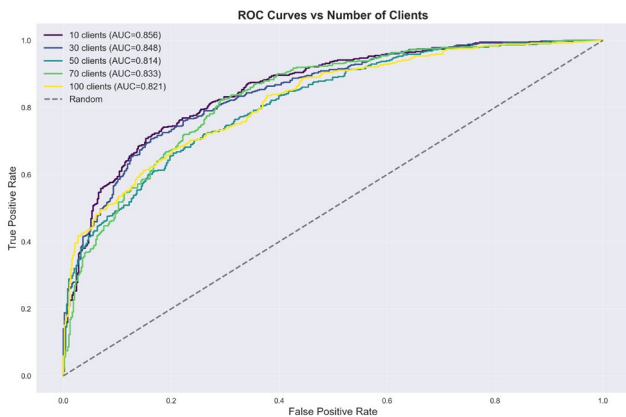

*Figure 2. ROC curves for different numbers of clients (10–100)*

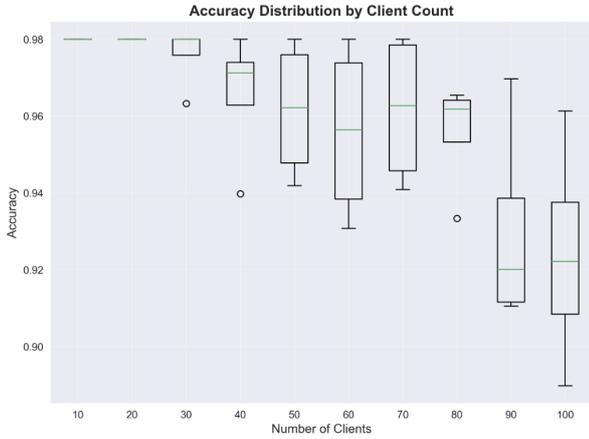

Figure 4. Accuracy distribution across varying client counts

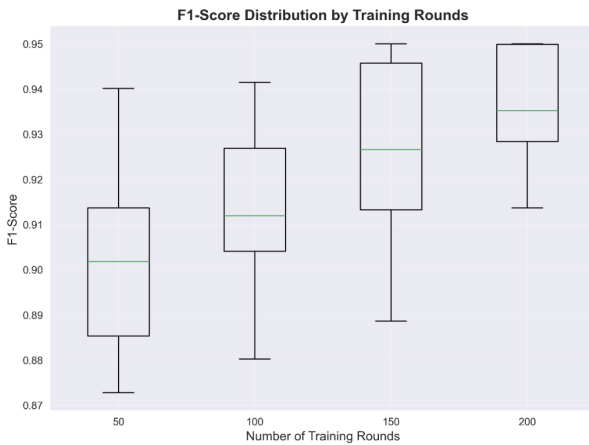

Figure 5. F1-Score distribution over different training rounds

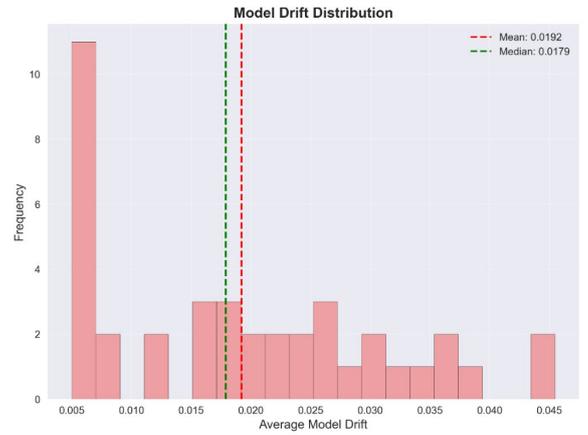

Figure 6. Distribution of average model drift across clients

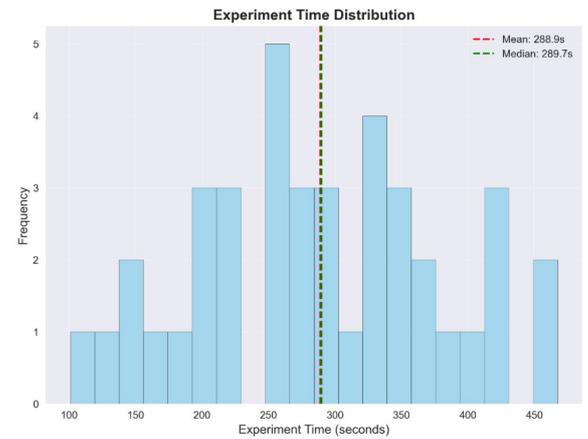

Figure 7. Distribution of experiment times indicating consistent and stable computational performance

Figure 6 illustrates the distribution of model drift, which serves as a key indicator of stability in federated learning. A substantial proportion of updates are concentrated at very low drift values (≈0.005–0.015), indicating that most clients remain well aligned with the global model. The mean drift (0.0192) and median drift (0.0179) are closely aligned, reflecting a balanced distribution of deviations across clients. Only a small subset of clients exhibits higher drift (up to ≈0.045), likely due to non-IID data distributions or lower data quality, which occasionally introduces noisier updates. Overall, the results demonstrate that the proposed framework maintains low average drift, ensuring stable convergence while effectively mitigating the risks of divergence across heterogeneous clients.

Finaly, figure 7 demonstrates the distribution of experiment times that most runs were concentrated around the 250–300 second range, with both the mean (288.9 s) and median (289.7 s) closely aligned. This proximity indicates that the experimental setup maintained consistent execution times across different trials, with minimal skewness or extreme outliers. Such stability in runtime reflects the efficiency of the proposed federated framework, ensuring predictable computational overhead even as the number of clients and training rounds varied.

Taken together, these results demonstrate that the proposed approach achieves state-of-the-art intrusion detection performance, while maintaining stability under non-IID conditions and ensuring scalability and efficiency in large-scale IoT deployments. Importantly, by addressing heterogeneity directly at the Mist layer through feature abstraction, and further stabilizing client updates with Fog-level FedProx aggregation, the framework overcomes one of the most critical obstacles in federated intrusion detection. This hierarchical strategy not only mitigates the negative effects of diverse sensing modalities and statistical imbalance but also ensures that the system remains both accurate and reliable across heterogeneous IoT environments.

### D. Comparing the proposed approach with other methods

Table 1 compares the proposed Mist-assisted federated learning approach with recent state-of-the-art intrusion detection methods evaluated on the ToN-IoT dataset. Abdel-Basset et al. [26] achieved moderate results with an accuracy of 94.85% and balanced precision–recall values around 93%. Booij et al. [27] reported a higher accuracy of 97.82% but suffered from a lower F1-score (92.12%), reflecting weaker precision–recall balance. Wang et al. [28] obtained more competitive results, with accuracy at 97.06% and F1-score of 96.94%. In contrast, the proposed Mist-based framework clearly outperforms prior work across all metrics, achieving 98.52% accuracy, 96.93% precision, 97.52% recall, and 96.93% F1-score. This improvement stems from the integration of Mist-level feature abstraction, which mitigates raw data heterogeneity before federated training, combined with utility-based client selection and Fog-level FedProx stabilization. These results confirm that embedding Mist intelligence within the FL pipeline significantly enhances robustness, scalability, and detection capability in heterogeneous IoT environments.

TABLE I.  COMPARISON OF THE PROPOSED APPROACH WITH OTHER RELATED WORKS

| Research | Accuracy (%) | F1-score (%) | Recall (%) | PRECISION (%) |
|---|---|---|---|---|
| Abdel et al. [26] | 94.85 | 93.13 | 93.09 | 93.17 |
| Booij et al. [27] | 97.82 | 92.12 | - | - |
| Wang et al. [28] | 97.06 | 96.94 | 96.67 | 97.23 |
| **Proposed Method** | **98.52** | **96.93** | **97.52** | **98.33** |

## V. CONCLUSION

This paper introduced a Mist-assisted federated learning framework tailored for intrusion detection in heterogeneous IoT environments. The framework addresses data heterogeneity directly at the Mist layer through feature abstraction, ensuring raw multi-modal inputs are standardized before training. In combination with utility-based client selection at the Edge, FedProx-based stabilization at the Fog, and global orchestration at the Cloud, the architecture overcomes critical limitations of conventional FL-based intrusion detection approaches. Comprehensive experiments on the TON-IoT dataset confirmed that the framework achieves state-of-the-art results, with detection accuracy consistently reaching 98–99%. Beyond raw performance, the system demonstrated stable convergence, low model drift, and predictable runtime efficiency, even under large-scale and non-IID conditions. The hierarchical design also reduces communication overhead and preserves privacy by ensuring that raw data remain local to IoT devices. This study opens multiple avenues for further exploration. Adaptive utility-based selection policies could dynamically adjust to client availability and reliability. Incorporating TinyML-powered anomaly detectors at the Mist would enable even more proactive defenses with minimal resource consumption. Additionally, evaluating the framework under adversarial or poisoning attacks is essential to strengthen resilience in hostile environments. Together, these directions highlight the potential of extending Mist-assisted FL into a scalable and future-proof paradigm for securing next-generation IoT ecosystems.

## REFRENCES